# An approximate method for controlling solid elastic waves by transformation media

Jin Hu[1], Zheng Chang[2] and Gengkai Hu[2*]

1 School of Information and Electronics, Beijing Institute of Technology, Beijing 100081, People's Republic of China.

2 School of Aerospace Engineering, Beijing Institute of Technology, Beijing 100081, People's Republic of China.

**Abstract**: By idealizing a general mapping as a series of local affine ones, we derive approximately transformed material parameters necessary to control solid elastic waves within classical elasticity theory. The transformed elastic moduli are symmetric, and can be used with Navier's equation to manipulate elastic waves. It is shown numerically that the method can provide a powerful tool to control elastic waves in solids in case of high frequency or small material gradient. Potential applications can be anticipated in nondestructive testing, structure impact protection, petroleum exploration and seismology.

[*] Corresponding author: hugeng@bit.edu.cn

Transformation method [1-3] provides an efficient tool to find necessary material distribution when wave pattern is prescribed in a space. Many interesting devices have been proposed especially with help of electromagnetic or acoustic metamaterials [4-8]. The method is also extended to control heat conduction [9] and matter waves [10]. These exotic properties are bestowed by form-invariance of Maxwell or Helmholtz equations under a general spatial transformation. For elastic waves in solid materials, Milton et al. [11] show that Navier's equation is transformed to Willis' one for a general spatial transformation, therefore the machinery developed for electromagnetic or acoustic waves cannot be directly applied to solid elastic waves. Due to the complicated nature of Willis' equation and its unusual constitutive equation, efforts are continually made to find approximate methods to control elastic wave in some degrees, since elastic waves are involved in broad engineering applications. For examples, in quasi-static limit, Zhou et al. [12] proposed an elastic cloak with impedance-matched method; Brun et al. [13] reported a cylindrical cloak for in-plane elastic wave with an asymmetric elasticity tensor; Farhat et al. [14] suggested a cloak for shielding thin plate bending wave in long wavelength limit. In this letter, we will idealize a general space transformation by local affine ones point by point, and propose an approximate method to control elastic waves. The approximation of the method will also be clarified in context of elastic ray theory.

When an initial space $\Omega$ is transformed to $\Omega'$ by a mapping $\mathbf{x}' = \mathbf{x}'(\mathbf{x})$, a physical process $F$ prescribed on the initial space with field quantity $\mathbf{u}$ and material parameter $\mathbf{C}$ is also transformed to the new space $\Omega'$ with new field quantity $\mathbf{u}'$ and material parameter $\mathbf{C}'$. The transformed relation between $\mathbf{u}'$ and $\mathbf{u}$, $\mathbf{C}'$ and $\mathbf{C}$ can be derived if the governing equation of the physical process is globally form-invariant. For elastic waves governed by Navier's equation, we will idealize a general transformation by a series of local affine ones point by point. Since Navier's equation locally keeps its form under an affine transformation [15], we will derive the transformed material parameters in this circumstance. To this end, we establish at any point $\mathbf{x}$ of $\Omega$ a local Cartesian frame $\mathbf{e}_i$ and at the point $\mathbf{x}' = \mathbf{x}'(\mathbf{x})$ of $\Omega'$ another local Cartesian frame $\mathbf{e}'_i$, uniquely determined by the mapping. With the assumption of local affine transformation, the governing equation will retain locally its form in the two spaces with respect to the two local Cartesian systems, i.e.

$$F(\mathbf{x}, \mathbf{u}, \mathbf{C}, t) = 0, \quad \mathbf{x} \in \Omega. \tag{1}$$

$$F(\mathbf{x}', \mathbf{u}', \mathbf{C}', t) = 0, \quad \mathbf{x}' \in \Omega', \tag{2}$$

Equation (2) imposes a constraint condition on $\mathbf{u}'$ and $\mathbf{C}'$. In addition, we assume that each type of energy is conserved at every point during the mapping, this will lead to another constraint condition, these constraint conditions will determine directly the transformed relations for $\mathbf{u}'$ and $\mathbf{C}'$. During a mapping, the space will experience locally a rigid rotation and a stretch operation, i.e., $\mathbf{A} = \nabla_\mathbf{x} \mathbf{x}' = \mathbf{V}\mathbf{R}$ [16], where $\mathbf{R}$ is rigid rotation, $\mathbf{V}$ is a stretch operation, and it can be expressed by its eigenvector $\mathbf{e}'_i$ and eigenvalues $\lambda_i$ as $\mathbf{V} = \lambda_1 \mathbf{e}'_1 \otimes \mathbf{e}'_1 + \lambda_2 \mathbf{e}'_2 \otimes \mathbf{e}'_2 + \lambda_3 \mathbf{e}'_3 \otimes \mathbf{e}'_3$. The $\mathbf{e}'_i$ establish a local Cartesian frame at each point in the transformed space $\Omega'$. Any physical quantity $\mathbf{q}$ in the initial space will experience with the space element the same rigid rotation, then stretch operation to reach $\mathbf{q}'$ in the transformed space, i.e., $\mathbf{V}_\mathbf{q} \mathbf{R} : \mathbf{q} \rightarrow \mathbf{q}'$. The transformed physical quantities must rescale themselves to satisfy Eq. (2) and the energy conservation condition; this provides a way to determine them. This general idea has been applied to electromagnetic and acoustic waves [17]. In the following, we will apply the above idea to elastodynamics, namely Navier's equation

$$\nabla \cdot \boldsymbol{\sigma} = \rho \frac{\partial^2 \mathbf{u}}{\partial t^2}, \quad \boldsymbol{\sigma} = \mathbf{C} : \nabla \mathbf{u}, \tag{3}$$

where $\mathbf{u}$ denotes displacement vector, $\boldsymbol{\sigma}$ is 2-order stress tensor, $\mathbf{C}$ is 4-order elasticity tensor and $\rho$ is density. Any transformed physical quantity $\mathbf{q}'$ can be written symbolically as

$$\mathbf{V_q R} : \mathbf{q} \to \mathbf{q}', \quad \mathbf{q} = \boldsymbol{\sigma}, \mathbf{u}, \mathbf{C}, \boldsymbol{\rho}, \tag{4}$$

where $\mathbf{R}$ establishes a local Cartesian frame $\mathbf{e}'_i$ from $\mathbf{e}_i$ (chosen arbitrary due to isotropy in the initial space) by a rotation, $\mathbf{V_q}$ has a diagonal form expressed in $\mathbf{e}'_i$ for the physical quantity $\mathbf{q}$, i.e.

$$\mathbf{V}_{\boldsymbol{\sigma}} = \mathrm{diag}[a_1, a_2, a_3], \mathbf{V}_{\mathbf{u}} = \mathrm{diag}[b_1, b_2, b_3], \mathbf{V}_{\mathbf{C}} = \mathrm{diag}[c_1, c_2, c_3], \mathbf{V}_{\boldsymbol{\rho}} = \mathrm{diag}[d_1, d_2, d_3], \tag{5}$$

$a_i, b_i, c_i$ and $d_i$ are scaling factors respectively for stress, displacement, modulus and density, they will be determined with help of the constraint conditions. During a rigid rotation, any attached physical quantity together with the frame $\mathbf{e}_i$ are rotated to $\mathbf{e}'_i$, so the component of the physical quantity in the local frame $\mathbf{e}'_i$ will not be altered by this rigid rotation. In the local Cartesian frame $\mathbf{e}'_i$, $\mathbf{V_q}$ is of diagonal form, so the transformed relations for the physical quantities from the frame $\mathbf{e}_i$ to the frame $\mathbf{e}'_i$ at each point can be written as [18]

$$\sigma'_{ij} = a_I a_J \sigma_{ij}, \quad u'_i = b_I u_i, \quad C'_{ijkl} = c_I c_J c_K c_L C_{ijkl}, \quad \rho'_{ij} = d_I \delta_{ij} \rho, \tag{6}$$

where $\delta_{ij}$ is Kronecker delta, the capital letter in index means the same value as its lower case but without summation.

Equation (3) is written in the initial local Cartesian frame $\mathbf{e}_i$ as

$$\frac{\partial \sigma_{ij}}{\partial x_i} = \rho \frac{\partial^2 u_j}{\partial t^2}, \quad \sigma_{ij} = C_{ijkl} \frac{\partial u_k}{\partial x_l}. \tag{7}$$

After a local affine transformation, this equation retains its form in the local Cartesian system $\mathbf{e}'_i$

in the transformed space $\Omega'$

$$\frac{\partial \sigma'_{ij}}{\partial x'_i} = \rho'_{ij} \frac{\partial^2 u'_i}{\partial t^2}, \quad \sigma'_{ij} = C'_{ijkl} \frac{\partial u'_k}{\partial x'_l}. \tag{8}$$

With help of Eqs. (6) and (8), and the differential relation between the two spaces $\partial/\partial x'_i = \partial/\lambda_i \partial x_i$ [17], we have

$$\frac{a_I a_J}{\lambda_I} \frac{\partial \sigma_{ij}}{\partial x_i} = d_J b_J \rho \frac{\partial^2 u_j}{\partial t^2}, \quad a_I a_J \sigma_{ij} = c_I c_J c_K c_L \frac{b_K}{\lambda_L} C_{ijkl} \frac{\partial u_k}{\partial x_l}. \tag{9}$$

Here the scaling factors are uniform at the considered point and its neighborhood respectively, due to the assumption of the local affine transformation. The conservations of strain and kinetic energies during the mapping give the following constraint for the scaling factors

$$\rho \delta_{ij} \frac{\partial u_i}{\partial t} \frac{\partial u_j}{\partial t} = \lambda_1 \lambda_2 \lambda_3 d_I b_I^2 \rho \left(\frac{\partial u_i}{\partial t}\right)^2, \quad \sigma_{ij}\left(\frac{\partial u_i}{\partial x_j} + \frac{\partial u_j}{\partial x_i}\right) = \lambda_1 \lambda_2 \lambda_3 a_I a_J \sigma_{ij} \left(\frac{b_I}{\lambda_J} \frac{\partial u_i}{\partial x_j} + \frac{b_J}{\lambda_I} \frac{\partial u_j}{\partial x_i}\right). \tag{10}$$

Comparing Eq. (9) directly with (7), together with Eq. (10), the scaling factors are found to satisfy

$$\frac{a_i a_j}{d_j b_j} = \lambda_i, \quad \frac{a_I a_J}{c_I c_J c_K c_L b_k} = \frac{1}{\lambda_I}, \quad a_i a_j b_I = \frac{\lambda_j}{\lambda_1 \lambda_2 \lambda_3}. \tag{11}$$

$\lambda_i$ is known when a mapping is provided, so the scaling factors $a_i$, $b_i$, $c_i$ and $d_i$ can be related to $\lambda_i$. Generally, Eq. (11) has non-unique solution, if we set $\mathbf{u}' = (\mathbf{A}^T)^{-1}\mathbf{u}$ as in the reference [11], i.e. $b_i = 1/\lambda_i$, then the scaling factors are uniquely determined as

$$a_i = \frac{\lambda_i}{\sqrt{\det \mathbf{A}}}, \quad b_i = \frac{1}{\lambda_i}, \quad c_i = \frac{\lambda_i}{\sqrt[4]{\det \mathbf{A}}}, \quad d_i = \frac{\lambda_i^2}{\det \mathbf{A}}. \tag{12}$$

In the global frame, the transformed relations of $\boldsymbol{\sigma}'$, $\mathbf{u}'$ and $\boldsymbol{\rho}'$ are given respectively by

$$\boldsymbol{\sigma}' = \frac{\mathbf{A}\boldsymbol{\sigma}\mathbf{A}^T}{\det \mathbf{A}}, \quad \mathbf{u}' = (\mathbf{A}^T)^{-1}\mathbf{u}, \quad \boldsymbol{\rho}' = \frac{\mathbf{A}\boldsymbol{\rho}\mathbf{A}^T}{\det \mathbf{A}}, \tag{13a}$$

The transformed relation of **C'** is expressed in the local Cartesian frame $\mathbf{e}'_i$ as

$$C'_{ijkl} = \frac{\lambda_I \lambda_J \lambda_K \lambda_L}{\det \mathbf{A}} C_{ijkl} . \qquad (13b)$$

Equation (13) provides the transformed material parameters for controlling solid elastic waves by transformation method. The transformed moduli possess necessary symmetry $C'_{ijkl} = C'_{ijlk} = C'_{jikl} = C'_{klij}$, as required in classical elasticity theory. If the material in the initial space is fluid, i.e., $C_{ijkl} = \kappa \delta_{ij} \delta_{kl}$, and $\mathbf{u}' = (\mathbf{A}^{\mathrm{T}})^{-1} \mathbf{u}$, both $\boldsymbol{\rho}'$ and $\mathbf{C}'$ agree with the results given in Ref. [19] for generalized acoustic waves obtained from a completely different method. If the materials in the initial and transformed spaces are both fluid, i.e., $\sigma_{ij} = P \delta_{ij}$, $\sigma'_{ij} = P' \delta_{ij}$, $C_{ijkl} = \kappa \delta_{ij} \delta_{kl}$ and $C'_{ijkl} = \kappa' \delta_{ij} \delta_{kl}$, the transformed relation given by the references [7,8,20] for acoustic wave can also be recovered.

To illustrate the proposed method, we design in the following an elastic rotator with the transformed material parameters given by Eq. (13), and validate them through numerical simulation. We propose the following mapping for the rotator

$$r' = r, \quad \theta' = \theta + f(r)\theta_0 , \qquad (14)$$

where $\theta_0$ is a rotation angle, $a$ and $b$ are the radii of inner and outer boundaries of the rotator, respectively, and $f(r)$ is a function used to satisfy the impedance-matched condition at the boundary. To this end, quintic polynomial $f(r)$ is used with $f(a)=1$, and $f(b)=f'(a)=f'(b)=f''(a)=f''(b)=0$ at the boundaries. Once the transformation is complete, the material parameters for the elastic rotator are obtained from Eq. (13), the constructed rotator is then simulated with COMSOL Multiphysics. We set $a$=0.1m, $b$=0.35m, the normalized Lame constants of the background are $\lambda = 2.3$, $\mu = 1$, and density is $\rho = 1$ with respect to fused silica [13]. A small circle of radius $r=0.01$m, a distance away from the rotator, can emit harmonic P-wave or S-wave respectively, it in turn impinges on the rotator. The PML conditions are imposed on the boundary of the computed region [21].

In order to quantify the approximation of the proposed transformed relation, we define a parameter $L = \max_{\Omega',i,j,k,l} |\nabla C'_{ijkl}| \gamma / C'_{ijkl}$, where $\gamma$ is the wavelength in the initial space, this parameter represents approximately the maximum material variation within a wavelength. It is expected that the proposed method will give better results when $L$ is small, i.e. more approaching the condition necessary for local affine transformation. Fig.1 shows the distributions of total displacement of the designed rotators with different $\theta_0$ and different incident frequencies $\gamma$ respectively, all are with an impinging S-wave. It is seen that the wave patterns are all rotated, as expected. The wave pattern is clearer in the case where $L$ is smaller, as shown in Fig.1. The wavefront becomes a little blurred when $L$ is large, as illustrated in Fig. 1(a). The same effect can also be observed for P-waves.

The local affine transformation implies that the material is considered locally to be homogeneous round a point (piece-wise constant), this assumption is commonly used in elastic rays theory [22]. According to this theory, for a wave of form $\mathbf{u}(\mathbf{x},t) = \mathbf{U}(\mathbf{x})F(t-T(\mathbf{x}))$, Navier's equation is reduced into the following two equations in high frequency approximation (elastic rays theory)

$$C_{ijkl} \frac{\partial T}{\partial x_l} \frac{\partial T}{\partial x_j} U_k - \rho_{ij} U_j = 0, \tag{15a}$$

$$C_{ijkl} \frac{\partial T}{\partial x_j} \frac{\partial U_k}{\partial x_l} + \frac{\partial}{\partial x_j}(C_{ijkl} \frac{\partial T}{\partial x_l} U_k) = 0. \tag{15b}$$

Equation (15a) is called eikonal equation describing ray path, and (15b) is transport equation characterizing wave amplitude along the ray. It can be shown that the transformed relation (13) verifies the form-invariant of eikonal equation, but keep transport equation (15b) form-invariant only if the scaling factors change slowly or their derivatives can be neglected, contrary to transformation optics and acoustics cases. This implies that the transformed material parameters can well control the wave path, but leave the wave amplitude to be controlled approximately. For very high frequency, eikonal equation overwhelms, the proposed method will lead to the nearly perfect results. The detail discusses of transformation method based on elastic rays theory will be presented elsewhere. As shown by Milton et al. [11], Navier's equation is transformed to Willis' equation, when material gradient is small or frequency is high, Willis' equation can be reduced to Navier's equation, and the transformation relation (13) can also be found from their theory. It must be stressed that the proposed method is different from the transformation methods existing in literatures, the transformed relations are derived directly from the constraint conditions, none of transformed relation is pre-assumed. As a result, the proposed method can offer more flexibility in material design, such as the impedance-matched condition in isotropic transformed materials [23].

In summary, we have proposed an approximate method to derive the transformed relation for controlling solid elastic wave. The method idealizes a general transformation by local affine ones, the local form-invariant of Navier's equation and the conservation of energy provide the constraint conditions to derive the transformed relations. It is shown numerically that the obtained transformed relation can be used within classical elasticity theory to control elastic waves, especially for the regime of high frequency or the small material gradient. It is also shown that the transformed material parameters control well ray path, but only approximately control the amplitude along the ray. Potential applications of the proposed results can be anticipated in nondestructive testing, structure impact protection, petroleum exploration and seismology [22,24].

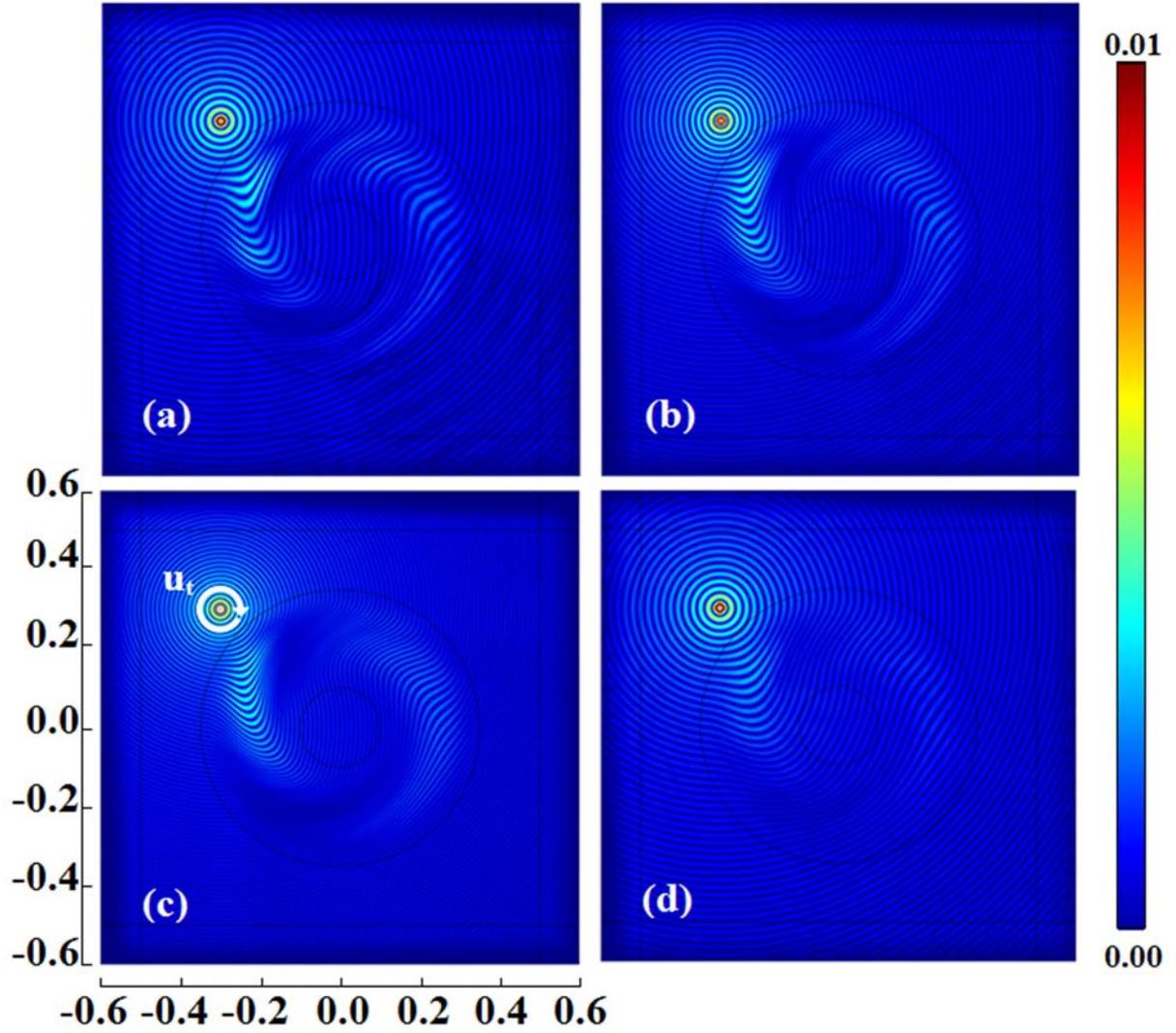

Figure 1. Simulation of rotator for elastic waves, total displacement field ($\sqrt{u_x^2 + u_y^2}$) of the generated S-wave (a) $\theta_0 = \pi/4$, $\gamma = 0.0375$m, $L \approx 1.748$, (b) $\theta_0 = \pi/4$, $\gamma = 0.03$m, $L \approx 1.398$, (c) $\theta_0 = \pi/4$, $\gamma = 0.0225$m, $L \approx 1.049$, (d) $\theta_0 = \pi/8$, $\gamma = 0.0375$m, $L \approx 0.774$.

**Acknowledgments:** The authors would like to thank Dr. TZ Li for helpful discussion on differential geometry theory, this work was supported by the National Natural Science Foundation of China (10832002), and the National Basic Research Program of China (2006CB601204).